\documentclass[runningheads]{llncs}
\usepackage{graphicx}

\usepackage[utf8]{inputenc}
\usepackage{cite}
\usepackage[T1]{fontenc}
\usepackage{lmodern}
\usepackage{textcomp}
\usepackage{latexsym}
\usepackage{amsmath}
\usepackage{amsfonts}
\usepackage{bm}
\usepackage{multirow}
\usepackage{tabularx}
\title{2D/3D Deep Image Registration by Learning \\3D Displacement Fields for Abdominal Organs}

\titlerunning{2D/3D Deep Image Registration by Learning 3D Displacement Fields}

\author{Ryuto Miura\inst{1} \and Megumi Nakao\inst{1} \and Mitsuhiro Nakamura\inst{2} \and Tetsuya Matsuda\inst{1}}

\institute{Graduate School of Informatics, Kyoto University, Kyoto, Japan.
\email{r-miura@sys.i.kyoto-u.ac.jp} \and Graduate School of Medicine, Kyoto University, Kyoto, Japan.}

\begin{document}
\maketitle
\begin{abstract}
Deformable registration of two-dimensional/three-dimensional (2D/3D) images of abdominal organs is a complicated task because the abdominal organs deform significantly and their contours are not detected in two-dimensional X-ray images. We propose a supervised deep learning framework that achieves 2D/3D deformable image registration between 3D volumes and single-viewpoint 2D projected images. The proposed method learns the translation from the target 2D projection images and the initial 3D volume to 3D displacement fields. In experiments, we registered 3D-computed tomography (CT) volumes to digitally reconstructed radiographs generated from abdominal 4D-CT volumes. For validation, we used 4D-CT volumes of 35 cases and confirmed that the 3D-CT volumes reflecting the nonlinear and local respiratory organ displacement were reconstructed. The proposed method demonstrate the compatible performance to the conventional methods with a dice similarity coefficient of 91.6 \% for the liver region and 85.9 \% for the stomach region, while estimating a significantly more accurate CT values.

\end{abstract}

\section{Introduction}

Three-dimensional (3D) medical imaging, such as computed tomography (CT) and magnetic resonance imaging (MRI), is widely used for diagnosis and pre-treatment or pre-operative planning. However, 3D medical images are unavailable during surgery or radiotherapy because of the limitations of imaging devices. Therefore, treatment is performed using only local two-dimensional (2D) images, such as endoscopic or X-ray images. Particularly in radiotherapy, organs and tumors can displace and deform during treatment, causing a risk to organs if the accuracy of irradiation cannot be guaranteed. In this study, we focus on 2D/3D image registration between pre-treatment 3D volumes and single-viewpoint 2D projection images, reconstructing 3D volumes during treatment.

2D/3D image registration has been widely investigated, especially for rigid bodies and those with minimal deformation such as skeletal structures. In conventional optimization-based methods, registration is achieved by optimizing the transformation parameters so that the intensity-based or contour-based similarity measure between the projection image simulated from 3D-CT volume, referred to as digitally reconstructed radiograph (DRR), and the X-ray image is maximized\cite{saitekika1, saitekika2}. However, such optimization-based methods have drawbacks owing to long computation time and unstable convergence characteristics. To overcome these issues, several learning-based methods of estimating transformation parameters using convolutional neural networks (CNNs) have been proposed\cite{goutai1, goutai2, goutai3}. For example, Miao et al. proposed a method for hierarchically estimating six degrees of freedom rigid transformation parameters from projection images using CNN, and showed that fast and accurate 2D/3D registration is possible\cite{goutai1}. However, these 2D/3D rigid registration methods require the geometry of the estimated object to be known and cannot be applied to 2D/3D deformable registration of non-rigid bodies such as abdominal organs, which deform significantly and whose contours are hard to detect in low-contrast X-ray images. Some efforts have been made for 2D/3D deformable registration\cite{Wu, X2S, IGCN}. Nakao et al. proposed a framework for reconstructing an organ mesh shape from a single X-ray image by learning the deformation of an abdominal organ from the mean shape using a graph convolutional network\cite{X2S, IGCN}. However, images were not reconstructed in these studies, and the problem in using them in radiotherapy planning, which requires dose calculation, is still to be solved. Image-based 2D/3D deformable registration has not yet been achieved or investigated.

This study proposes a new learning-based 2D/3D deformable image registration method for abdominal organs, which reconstructs 3D-CT volume from a single-viewpoint X-ray image. Specifically, this study targets the patient respiration, which causes abdominal organs to displace or deform significantly and their contours cannot be detected in low-contrast 2D projection images. The proposed method achieves this by extending VoxelMorph\cite{voxelmorph}, which was proposed as a 3D/3D registration framework for brain MRI, introducing a 3D displacement field (3D-DF) generator from 3D-CT volume and DRR images. We evaluated the performance of the proposed method using 4D-CT volumes for one respiratory cycle of 35 patients who underwent intensity-modulated radiotherapy. 

This study is the first to achieve 2D/3D image registration on a single X-ray image targeted to the abdominal organs. With the proposed method, for example, the 3D-CT volume reflecting displacements of organs at risk or tumor volumes become available from only X-ray image taken during treatment, and that leads to adaptive radiotherapy.

\section{Method}
\subsection{Dataset and Problem Definition}
\begin{figure}[t]
 \centering
 \includegraphics[width=12cm]{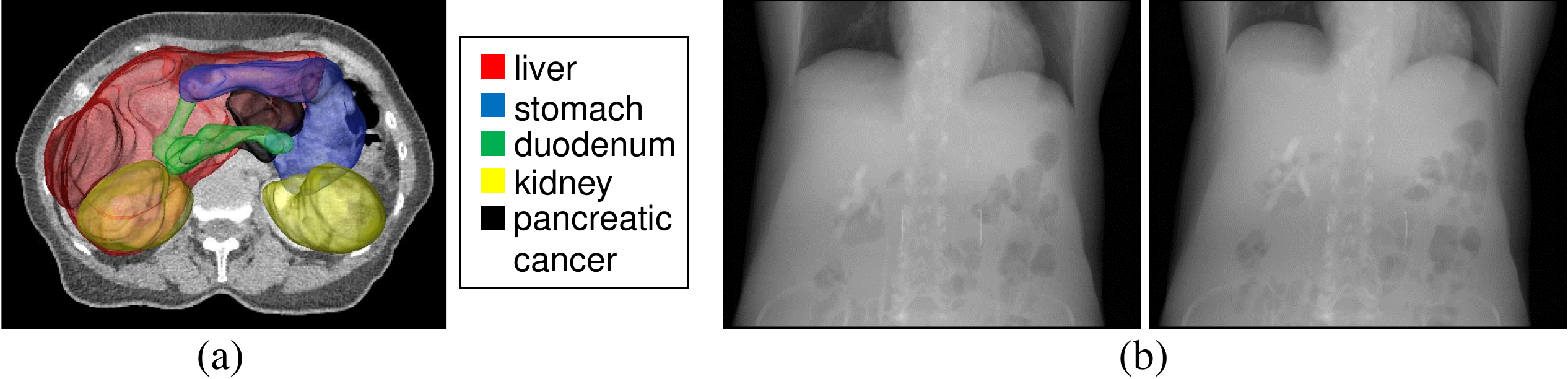}
 \caption{Experimental dataset, (a) 3D-CT slice image with 3D shapes of organs, (b) Digitally reconstructed radiographs (DRRs) at $t = 0$ of end-inhalation (left figure) and at $t = 50$ of end-exhalation (right figure). Displacement of anatomical structures due to patient respiration can be seen.}
 \label{fig:dataset}
\end{figure}

We used abdominal 4D-CT volumes of 35 patients who underwent intensity-modulated radiotherapy for pancreatic cancer. Each 4D-CT volume consisted of 10 time phases ($t = 0, 10,\cdots,90 \,\%$) of 3D-CT volumes for one respiratory cycle, with $t = 0$ and $t = 50$ corresponding to the end-inhalation and end-exhalation phases, respectively. Each 3D-CT volume consisted of 88-152 slices (voxel resolution: $1.0\,\mathrm{mm} \times 1.0\,\mathrm{mm} \times 2.5\,\mathrm{mm}$) with $512 \times 512$ pixels, and the regions of organs such as the liver, stomach, duodenum, were manually labeled by board-certified radiation oncologists\cite{SDM}, as shown in Fig.\ref{fig:dataset}(a). Each front-view DRR (Fig.\ref{fig:dataset}(b)) was generated from the corresponding 3D-CT volume, and each slice of 3D-CT volume and DRR were resampled to $128 \times 128$ pixels with isotropic pixels.

The aim of this study is to find the local correspondence between DRR and 3D-CT volume of the abdominal region taken at different time phases during respiratory deformation in a patient. We use the 3D-CT volume as the source and DRR as the target and generate the 3D-DF at the voxel level of the 3D-CT volume based on the image intensity to perform 2D/3D deformable image registration from the source to the target and evaluate the performance. The data used in this study have characteristics of large displacement and nonlinear deformation of organs in each respiratory cycle, and the contours of each organ are unclear in low-contrast DRR. Therefore, the performance of registration was not revealed even in the previous image registration studies\cite{voxelmorph, registration review}. In this study, to confirm the relationship between organ deformation due to respiration and the registration performance of the proposed method, the time phase of 3D-CT volume was fixed at $t = 0$ (end-inhalation), and that of the target DRR was fixed at $t = T$, where $T$ was varied from $10$ to $90$. In the proposed method, the target image and source volume are concatenated and input to the network as a $128 \times 128 \times 128 \times 2\,\mathrm{channel}$ 3D volume. The first channel is a 3D-CT volume, and the second channel contains only the pixel information of two DRR images ($t = 0, T$) after initializing all voxel values to 0. To train the network, 7-fold cross-validation was adopted by dividing 35 patients into 7 groups of 5 patients each. Training was performed using 4D-CT volumes of 30 patients (300 3D-CT volumes), and the 4D-CT volumes of the other 5 patients were used for testing.

\subsection{Proposed Framework}
\begin{figure}[t]
 \centering
 \includegraphics[width=12cm]{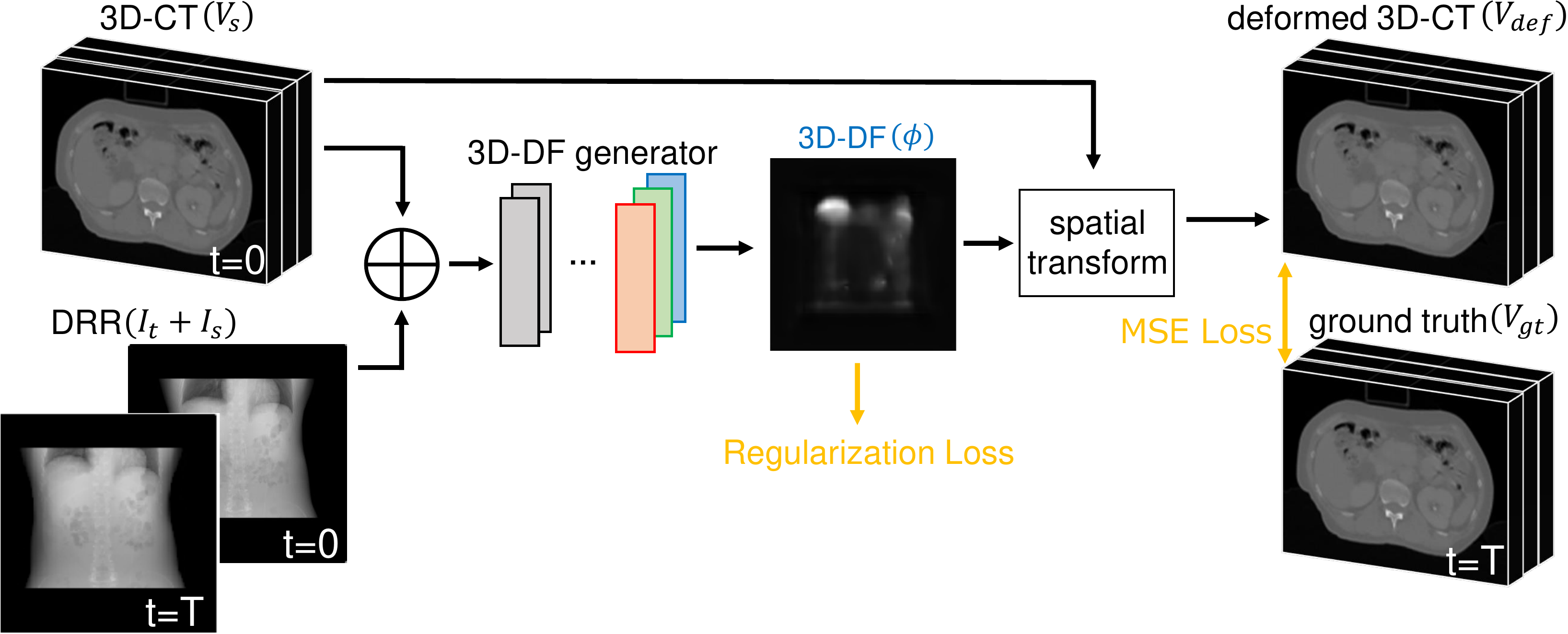}
 \caption{Overview of the proposed framework. The network learns the translation from the 2-channel 3D volume, consisted of a 3D-CT volume and DRR images, to the 3-channel 3D dispplacement field (3D-DF). $\oplus$ represents the concatenation of a source 3D-CT volume and target volume consists of DRR images.}
 \label{fig:framework}
\end{figure}

Fig.\ref{fig:framework} shows the framework of the proposed method. A source 3D-CT volume at $t=0$ ($V_s$), a DRR at the same time phase ($I_s$), and a target DRR at $t=T$ ($I_t$) are concatenated and input into the network, and the CNN-based 3D-DF generator learns the displacement of the anatomical structure. Using the displacement vectors of each voxel, $V_s$ is spatially transformed, and the intensity-based similarity between the reconstructed 3D-CT volume ($V_{def}$) and ground truth 3D-CT volume at $t=T$ ($V_{gt}$) is measured, and the weights of the network are optimized. We describe each module of the proposed framework in detail in this section.

\begin{figure}[t]
 \centering
 \includegraphics[width=7cm]{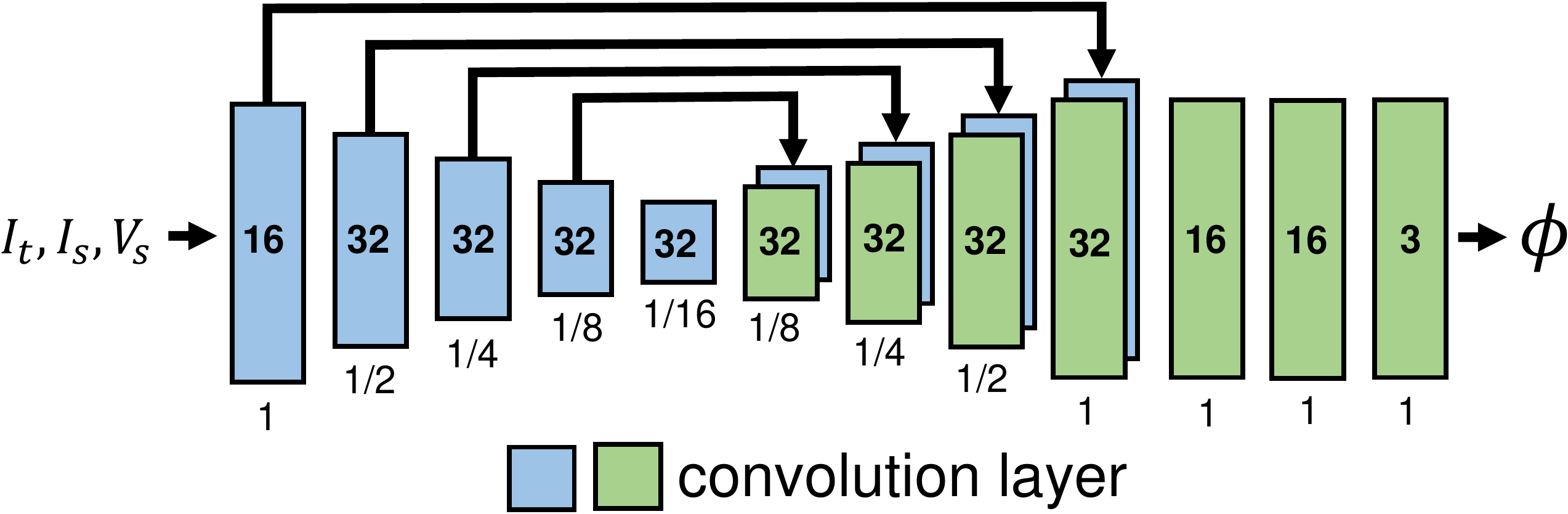}
 \caption{Architecture of the 3D-DF generator implemented using 3D U-net. Each rectangle represents a 3D convolution layer. The size of each layer with respect to its input and the number of channels of each layer are marked below and inside the rectangle, respectively.}
 \label{fig:CNN}
\end{figure}

\subsubsection{3D-DF Generator}
We describe the 3D-DF generator architecture that learns the translation from input images to 3D-DF. For the target DRR image, $I_s$ is given as input information in addition to $I_t$ to extract displacement features from the differences in pixel values between the DRR images. Therefore, the generator takes an input formed by concatenating $V_s$, $I_s$ and $I_t$ into a 2-channel 3D volume. The displacement vector field $\mathbf{u}$ is calculated by the 3D-DF generator, and the 3-channel 3D-DF $\phi = Id+\mathbf{u}$ is formed using identity transform ($Id$) and $\mathbf{u}$, where the displacement vector is stored in each voxel of $V_s$. In the proposed method, the generator function is implemented using 3D U-net\cite{U-net}, which is a CNN-based network with encoder-decoder sections. Fig.\ref{fig:CNN} shows the structure. Each layer of the encoder and decoder stage performs 3D convolutions with a kernel size of 3, and a stride of 2, and the activation function after each convolution is LeakyReLU with parameter 0.2. Each layer of the decoder is followed by an upsampling layer to bring the volume back to its original resolution. In addition, 3D U-Net propagates the features learned in the encoder stage to the decoder stage by concatenating the corresponding feature maps in each layer of the encoder and decoder stages, achieving precise and anatomically registered volume generation.

\subsubsection{Spatial Transform of 3D-CT using 3D-DF}
In the proposed method, 3D-CT volume $V_{def}$ is reconstructed by spatial transformation of $V_s$ using a 3D-DF $\phi$, while preserving the CT values. The optimal parameter values are learned by minimizing the differences between $V_{def}$ and $V_{gt}$. To use the standard gradient-based optimization method, we compute the value of each voxel in $V_{def}$ using differentiable operations\cite{spatial transformer}. \\
First, for each voxel $\mathbf{p}$, we calculate its corresponding location $\mathbf{p}^{\prime}=\mathbf{p}+\mathbf{u}(\mathbf{p})$ in $V_s$. $\mathbf{p}^{\prime}$ is often not at integer location since $\mathbf{u}$ is a continuous function; however, voxel values are defined only at integer locations in $V_s$. Therefore, the value of $V_{def}(\mathbf{p})$ is linearly interpolated by the values of the eight voxels adjacent to $\mathbf{p}^{\prime}$, as shown in Eq.\ref{eq:sp}.

\begin{equation}
\label{eq:sp}
V_{def}(\mathbf{p})=\sum_{\mathbf{q} \in \mathcal{Z}\left(\mathbf{p}^{\prime}\right)} V_{s}(\mathbf{q}) \prod_{d \in\{x, y, z\}}\left(1-\left|\mathbf{p}_{d}^{\prime}-\mathbf{q}_{d}\right|\right),
\end{equation}
where $\mathcal{Z}\left(\mathbf{p}^{\prime}\right)$ are the voxels adjacent to $\mathbf{p}^{\prime}$, weighted by each distance from $\mathbf{p}^{\prime}$ and summed. This allows us to compute the gradients or subgradients, which in turn allows for error back propagation during optimization.

\subsubsection{Loss Functions}
To obtain a $V_{def}$ with high similarity to $V_{gt}$ and smooth deformation, we introduce two loss functions: mean squared error (MSE) loss $\mathcal{L}_{MSE}$ and regularization loss $\mathcal{L}_{\text {smooth}}$. $\mathcal{L}_{MSE}$ imposes constraints on the differences in voxel values and $\mathcal{L}_{\text {smooth}}$ imposes constraints on the local displacement of $\phi$. First,$\mathcal{L}_{MSE}$ evaluates the error in voxel values of the corresponding voxels between $V_{def}$ and $V_{gt}$. $\mathcal{L}_{MSE}$ is defined by

\begin{equation}
\label{eq:MSE}
\mathcal{L}_{MSE}(V_{gt}, V_{def})=\frac{1}{|\Omega|} \sum_{\mathbf{p} \in \Omega}[V_{gt}(\mathbf{p})-[V_{def}](\mathbf{p})]^{2},
\end{equation}
where $\Omega$ is the set of all voxels in the 3D-CT volume. This loss function induces the convergence of each voxel in $V_{def}$ to the corresponding voxel location in $V_{gt}$.\\
Minimizing the $\mathcal{L}_{MSE}$ brings $V_{def}$ similar to $V_{gt}$, but may generate an unsmooth $\phi$ that is physically impossible. Therefore, we introduce $\mathcal{L}_{\text {smooth}}$ to regularize the gradient of the 3D-DF $\mathbf{u}$ and generate a smooth $\phi$, as follows.

\begin{equation}
\label{eq:smooth}
\mathcal{L}_{\text {smooth}}(\phi_{pre})=\sum_{\mathbf{p} \in \Omega}\|\nabla \mathbf{u}(\mathbf{p})\|^{2}.
\end{equation}

By using this loss function, the network can learn to obtain a $V_{def}$ that is similar to $V_{gt}$ while representing smooth deformations of organs.\\
The total loss is the weighted sum of three loss functions:

\begin{equation}
\label{eq:loss_total}
\mathcal{L}_{total}(V_{gt}, V_{def}, \phi_{gt}, \phi_{pre})=\mathcal{L}_{\text {MSE}}(V_{gt}, V_{def})+\lambda \mathcal{L}_{\text {smooth }}(\phi_{pre})+\gamma\mathcal{L}_{DVF}(\phi_{gt}, \phi_{pre})
\end{equation}

\begin{equation}
\label{eq:loss_DVF}
\mathcal{L}_{DVF}(\phi_{gt}, \phi_{pre})=\frac{1}{|\Omega|} \sum_{\mathbf{p} \in \Omega}[\mathbf{u}_{gt}(\mathbf{p})-\mathbf{u}_{pre}(\mathbf{p})]^{2}
\end{equation}

In the proposed method, we used 0.05 for $\lambda$, which is recommended by VoxelMorph\cite{voxelmorph}.

\section{Experiments}
In the experiments, we performed 2D/3D image registration between low-contrast DRR and 3D-CT volume obtained at different time phases in a patient to confirm the performance of reconstructing the 3D-CT volume reflecting the respiratory organ motion. In addition, the performance of the proposed method was compared with that of the conventional 2D/3D image registration methods. 

In this experiment, the mean absolute error (MAE) between the corresponding voxel values of the reconstructed 3D-CT volume and the ground truth 3D-CT volume, and the Dice similarity coefficient (DSC) of the organ volumes were used as the volume similarity metrics. MAE evaluates the average difference in CT values between the 3D-CT volumes, and DSC measures the volume overlap between the deformed and the ground-truth organs. The liver and stomach were employed as the target of the similarity evaluation by DSC, where the stomach may not have detectable features in the DRR while the liver have that (e.g., the diaphragm). In addition to these evaluation metrics, we visualized the reconstructed 3D-CT volumes and each 3D-DF, confirming the local deformations and estimated the volume quality and precisely evaluated the performance of the proposed method.

We compared the performance of the proposed method with a conventional optimization-based 2D/3D rigid registration method and a learning-based method based on the conventional 2D/3D registration by evaluating the similarity within 2D projection images only. In the former method, based on the existing methods\cite{saitekika1, saitekika2}, we implemented a framework in which the rigid transformation parameters were iteratively optimized to maximize the intensity-based similarity between DRR images. In the latter method, the 3D-CT volume was reconstructed using a 2D displacement field (2D-DF) only, excluding the displacement in the dorsoventral axis direction, obtained by 2D/2D deformable image registration between DRR images. Since the proposed method learns the 3D-DF from the input images, we used the comparison method to verify the performance of 3D-CT volume reconstruction by estimating the displacement in the dorsoventral direction, which is difficult to estimate from DRR alone. We implemented 2D/2D registration between DRR images by replacing the source 3D-CT volume with the DRR of $t = 0$ and the target volume and the ground-truth 3D-CT volume with the DRR of $t = T$ in the proposed method. The obtained 2D-DF was used to spatially transform all the coronal slices of the source 3D-CT volume, reconstructing the approximate 3D-CT volume. This comparison method, hereinafter, is referred to as 2D-DF.

The entire network was implemented using Python 3.8.10, Keras with a Tensorflow backend, and the Tensorflow-GPU library. The network was trained using an Adam optimizer with a learning rate of $1 \times 10^{-4}$. The batch size was set to 4, the number of epochs to 200, and it took 5 hours for each training in 7-fold cross-validation on a single NVIDIA TITAN RTX. For 2D-DF, the batch size was set to 32 and other conditions were the same as the proposed method.

\begin{figure}[t]
 \centering
 \includegraphics[width=12cm]{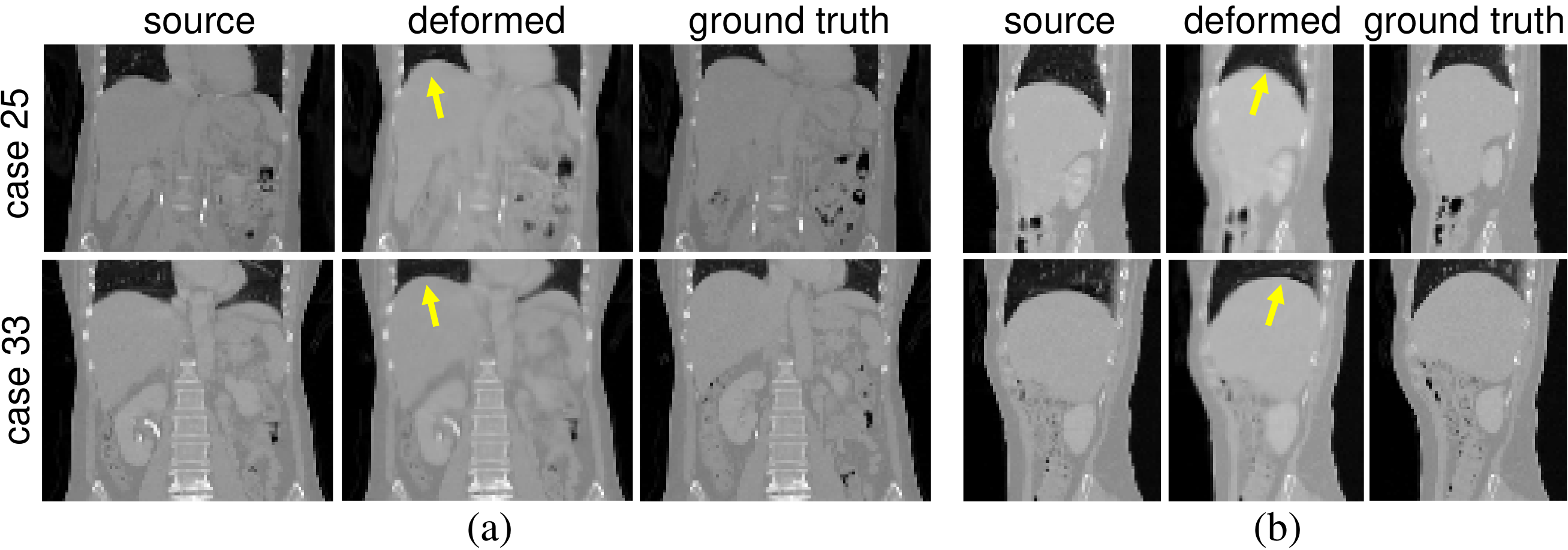}
 \caption{3D-CT volume reconstruction results of large respiratory organ motion cases, (a) coronal slices, (b) sagittal slices. The arrows indicate the diaphragm with obvious contour, and local and nonlinear deformations of the organ can be seen.}
 \label{fig:slices}
\end{figure}

\begin{table}[t]
 \caption{quantitative comparison of 3D-CT volume reconstruction performance, where Rigid Reg represents the optimization-based 2D/3D rigid registration. The mean±standard deviation of mean absolute error (MAE) and Dice similarity coefficient (DSC).}
 \label{table:quantitative}
 \centering
  \begin{tabular}{p{10em} p{6em} p{6em} p{6em} p{6em}}
   \hline
   & Initial & Rigid Reg & 2D-DF & Proposed \\
   \hline 
   MAE & $89.2\pm27.6$ & $248.8\pm80.1$ & $91.1\pm28.5$ & $69.4\pm19.7$ \\
   liver DSC [\%] & $84.6\pm6.9$ & $73.6\pm12.3$ & $90.7\pm3.5$ & $91.6\pm3.4$\\
   stomach DSC [\%] & $73.8\pm12.8$ & $51.1\pm21.6$ & $84.3\pm7.1$ & $85.9\pm4.7$ \\
   \hline
  \end{tabular}
\end{table}
The experiments were conducted with target DRR time phases ranging from $t=10$ to $t=90$, and the evaluation results for all 35 patients are listed in Table \ref{table:quantitative}, where the values of each evaluation metric were calculated using the results for $t=50$ because the respiratory organ deformation between two states of $t=0$ and $t=50$ causes the maximum estimation error in one respiratory cycle. The results show that the proposed method improves the values of all evaluation metrics compared to the initial state. In particular, DSC for the stomach, which have very few detectable features in the projected image, were improved by more than 10 \%. Although no significant difference was observed between the DSC of organs using the proposed method and 2D-DF, the MAE showed a significant difference (one-way analysis of variance, ANOVA; $p < 0.05$ significance level). Fig.\ref{fig:slices} shows the results of the test using the proposed method for two cases with large organ displacement due to respiration from the viewpoint of MAE, where the target DRR time phase was set to $t=50$ (end-exhalation state, the largest organ displacement compared to the source state). In each coronal slice (Fig.\ref{fig:slices}(a)) and sagittal slice (Fig.\ref{fig:slices}(b)) of the deformed 3D-CT volume, smooth and localized organ deformations similar to the ground-truth slice can be observed, especially in the diaphragm area, which is easily visible.

\begin{figure}[t]
 \centering
 \includegraphics[width=12cm]{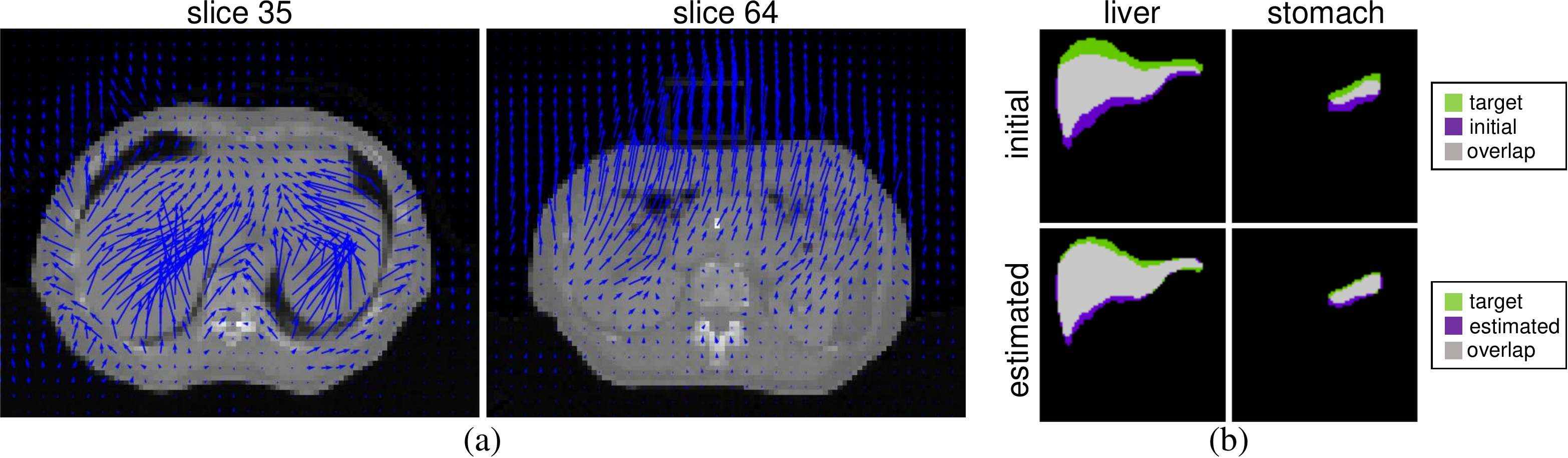}
 \caption{Visualization of 3D-DF and estimated organ regions, (a) 3D-DF in axial slices, (b) estimation results of liver and stomach regions.}
 \label{fig:DVF}
\end{figure}

Fig.\ref{fig:DVF}(a) shows the 3D-DF in the axial slices of the deformed 3D-CT volume estimated by the proposed method, where the target DRR time phase was set to $t=50$. The displacement vectors, represented by the blue arrows in the axial slices capturing different anatomical structures, demonstrate that the proposed method estimates the nonlinear and local organ displacements including in the dorsoventral direction. Fig.\ref{fig:DVF}(b) shows the reconstructed liver and stomach regions in a coronal slice, where the region of each deformed organ, the target, and the overlap is colored in purple, green and white, respectively. The estimated organ regions are closer to the target than in the initial state. On the other hand, there are estimation errors in the upper and lower parts of each organ region. This indicates that the proposed method may have a limitation in estimating the organ displacement in the body axis direction when the organ displacements are large.

\section{Conclusion}
This paper proposed a first deep learning framework to achieve deformable image registration of the 3D-CT volume with low-contrast 2D projection image targeted to the abdominal region. To estimate the nonlinear and local organ motions, we introduced a 3D-DF generator which extracts displacement features from the initial 3D-CT volume and the target 2D projection images, translating to 3D-DF. Compared to the conventional method, the proposed method demonstrated the improved performance in terms of CT value estimation required for dose calculation in radiotherapy plannings. Since the proposed method cannot estimate day-to-day specific anatomical changes such as gas generation during radiotherapy, our future work includes improvement of the framework to address this.



\begin{thebibliography}{99}

\bibitem{saitekika1}
C. Gendrin, H. Furtado, C. Weber, C. Bloch, M. Figl, S.A. Pawiro, H. Bergmann, M. Stock. G. Fichtinger, D. Georg, W. Birkfellner, "Monitoring tumor motion by real time 2D/3D registration during radiotherapy," Radiotherapy and Oncology, vol. 102, no. 2, pp. 274-280, 2012.

\bibitem{saitekika2}
C.J.F. Reyneke, M. Lüthi, V. Burdin, S.T. Douglas, T. Vetter, T.E. Mutsvangwa, "Review of 2-D/3-D Reconstruction Using Statistical Shape and Intensity Models and X-Ray Image Synthesis: Toward a Unified Framework," IEEE Reviews in Biomedical Engineering, vol. 12, pp. 269-286, 2018.

\bibitem{goutai1}
S. Miao, Z.J. Wang, R. Liao, "A CNN Regression Approach for Real-Time 2D/3D Registration," IEEE Trans. Medical Imaging, vol. 35, no. 5, pp. 1352-1363, 2016.

\bibitem{goutai2}
J. Zheng, S. Miao, R. Liao, "Learning CNNs with Pairwise Domain Adaption for Real-Time 6DoF Ultrasound Transducer Detection and Tracking from X-Ray Images," Medical Image Computing and Computer-Assisted Intervention (MICCAI), pp. 646-654, 2017.

\bibitem{goutai3}
A. Presenti, Z. Liang, L.F.A. Pereira, J. Sijbers, J. De Beenhouwer, "CNN-based Pose Estimation of Manufactured Objects During Inline X-ray Inspection," In 2021 IEEE 6th International Forum on Research and Technology for Society and Industry (RTSI), pp. 388-393, 2021.

\bibitem{Wu}
S. Wu, M. Nakao, J. Tokuno, T. Chen-Yoshikawa, and T. Matsuda, "Reconstructing 3D Lung Shape from a Single 2D Image during the Deaeration Deformation Process using Model-based Data Augmentation," IEEE Int. Conf. on Biomedical and Health Informatics (BHI), pp. 1-4, 2019.

\bibitem{X2S}
F. Tong, M. Nakao, S. Wu, M. Nakamura and T. Matsuda, "X-ray2Shape: Reconstruction of 3D Liver Shape from a Single 2D Projection Image", IEEE Eng Med and Biol Soc (EMBC), pp. 1608-1611, 2020.

\bibitem{IGCN}
M. Nakao, M. Nakamura, T. Matsuda, "Image-to-Graph Convolutional Network for Deformable Shape Reconstruction from a Single Projection Image," Medical Image Computing and Computer-Assisted Intervention (MICCAI), pp. 259-268, 2021.

\bibitem{voxelmorph}
G. Balakrishnan, A. Zhao, M.R. Sabuncu, J. Guttag, A.V. Dalca, "VoxelMorph: A Learning Framework for Deformable Medical Image Registration," IEEE trans. Medical Imaging, vol. 38, no. 8, pp. 1788-1800, 2019.

\bibitem{SDM}
M. Nakao, M. Nakamura, T. Mizowaki, T. Matsuda, "Statistical deformation reconstruction using multi-organ shape features for pancreatic cancer localization," Medical Image Analysis, Vol. 67, p. 101829, 2021.

\bibitem{registration review}
G. Haskins, U. Kruger, P. Yan, "Deep learning in medical image registration: a survey," Machine Vision and Applications, vol. 31, no. 1, pp. 1-18, 2020.

\bibitem{U-net}
O. Ronneberger, P. Fischer, T. Brox, "U-Net: Convolutional Networks for Biomedical Image Segmentation," Medical Image Computing and Computer-Assisted Intervention (MICCAI), pp. 234-241, 2015.

\bibitem{spatial transformer}
R. Bajcsy, S. Kovačič, "Multiresolution elastic matching," Computer Vision, Graphics, and Image Processing, vol. 46, no. 1, pp. 1-21, 1989.

\end{thebibliography}
\end{document}